\documentclass[12pt]{iopart}

\usepackage{iopams}

\begin{document}

\title{Bounds on large extra dimensions from photon fusion process in SN1987A}

\author{V H Satheeshkumar$^{1,2}$ and P K Suresh$^{1}$}

\address{$^{1}$ School of Physics, University of Hyderabad, Hyderabad 500 046, India.\\ $^{2}$Department of Physics, Sri Bhagawan Mahaveer Jain College of Engineering, Jain Global Campus, Kanakapura Road, Bangalore 562 112, India.}
\ead{vhsatheeshkumar@gmail.com, pkssp@uohyd.ernet.in}

\begin{abstract}
The constraint on the ADD model of extra dimensions coming from photon annihilation into Kaluza-Klein graviton in supernova cores is revisited. In the two photon process for a conservative choice of the core parameters, we obtain the bound on the fundamental Planck scale $M_* \gtrsim$  1.6 TeV. The combined energy loss rate due to nucleon-nucleon brehmstrahlung  and photon annihilation processes is rederived, which shows that the combined bounds  add  only  second decimal place to $M_*$. The present study can strengthen the results that are available in the current literature for the graviton emission from SN1987A which puts a very strong constraints on models with large extra dimensions for the case of $n=3$ .
\end{abstract}



\maketitle

\section{Introduction}
Stars are potential sources for weakly interacting particles such as neutrinos,
gravitons, axions, and other new particles that can be produced by nuclear
reactions or by thermal processes in the hot stellar interior.  The solar
neutrino flux is now routinely measured with such a precision that compelling
evidence for neutrino oscillations has accumulated. The measured neutrino burst
from supernova SN1987A has been used to derive many useful limits. Even when the
particle flux can not be measured directly, the absence of visible decay
products, notably x- or $\gamma$-rays, can provide important information. The
properties of stars themselves would change if they lose too much energy into a
new channel.  This ``energy-loss argument'' has been widely used to constrain a
long list of particles and there properties. All of this has been extensively
reviewed \cite{Raffelt,Raffelt:1999tx}

The  extra dimensional scenario due to Arkani-Hamed, Dimopoulos and Dvali (ADD)
\cite{ADD}, model predicts a variety of novel signals which can be tested using
table-top experiments, collider experiments, astrophysical or cosmological
observations. It has been pointed out that one of the strongest bounds on models of extra
dimensions comes from SN1987A \cite{ADDphen}. Various authors have done
calculations to place such constraints on the extra dimensions
\cite{Cullen:1999hc}-\cite{Hannestad:2003yd}.
In this paper, we calculate the energy loss rate due to graviton emission from
SN1987A by
photon-photon annihilation  and derive the bounds on extra
dimensions. We combine the result with that of nucleon-nucleon brehmstralung
process and derive the corresponding bound on large extra dimensions.

Physically, there are two fundamental types of supernovae (SNe), based on what
mechanism powers them: the thermonuclear SNe (Type I SNe) and the core-collapse ones (Type II SNe). 
The core-collapse SNe are the class of explosions which mark the
evolutionary end of massive stars ($M\gtrsim 8\,M_\odot$). Such stars have the
usual onion structure with several burning shells, an expanded envelope, and a
degenerate iron core that is essentially an iron white dwarf. The core mass
grows by the nuclear burning at its edge until it reaches the Chandrasekhar
limit. The collapse can not ignite nuclear fusion because iron is the most
tightly bound nucleus. Therefore, the collapse continues until the equation of
state stiffens by nucleon degeneracy pressure at about nuclear density
($3\times10^{14}\, {\rm gcm^{-3}}$).  At this ``bounce'' a shock wave forms,
moving outward and expelling the stellar mantle and envelope. The explosion is a
reversed implosion, the energy derives from gravity, not from nuclear energy.
 Within the expanding nebula, a compact
object remains in the form of a neutron star or perhaps sometimes a black hole.
The kinetic energy of the explosion carries about 1\% of the liberated
gravitational binding energy of about $3\times10^{53}\,{\rm erg}$, 99\% going
into neutrinos. This powerful
and detectable neutrino burst is the main astro-particle interest of
core-collapse SNe. In core-collapse SNe only $10^{-4}$ of the total energy
shows up as light, i.e. about 1\% of the kinetic explosion energy, hence they
are dimmer than SNe-Ia, and are not useful as standard candles.

In the case of SN1987A, about $10^{53}$ ergs of gravitational binding energy
was released in few seconds and the neutrino fluxes were measured by Kamiokande
\cite{Kamio} and IMB \cite{IMB} collaborations. Numerical neutrino light curves
can be compared with the SN1987A data where the measured energies are found to
be ``too low.''  For example,
the numerical simulation in \cite{Totani:1997vj} yields time-integrated values
$\langle E_{\nu_e}\rangle\approx13\,{\rm MeV}$, $\langle
E_{\bar\nu_e}\rangle\approx16\,{\rm MeV}$, and $\langle
E_{\nu_x}\rangle\approx23\,{\rm MeV}$.  On the other hand, the data imply
$\langle E_{\bar\nu_e}\rangle=7.5\,{\rm MeV}$ at Kamiokande and 11.1 MeV at
IMB \cite{Jegerlehner:1996kx}.  Even the 95\% confidence range for Kamiokande
implies $\langle E_{\bar\nu_e}\rangle<12\,{\rm MeV}$. Flavor oscillations would
increase the expected energies and
thus enhance the discrepancy \cite{Jegerlehner:1996kx}. It has remained unclear
if these and other anomalies of the SN1987A neutrino signal should be blamed on
small-number statistics, or point to a serious problem with the SN models or the
detectors, or is there a new physics happening in SNe?

Since we have these measurements already at our disposal, now if we propose some
novel channel through which the core of the supernova can lose energy, the
luminosity in this channel should be low enough to preserve the agreement of
neutrino observations with theory. That is,
\begin{equation}
{\cal L}_{new\, channel} \lesssim 10^{53}\, ergs\, s^{-1}.
\end{equation}
This idea was earlier used to put the strongest experimental upper bounds on the
axion mass \cite{axions}. Here, we consider the emission of the
higher-dimensional gravitons from the core. Once these particles are produced,
they can  escape into the extra dimensions, carrying energy away with them. The
constraint on the luminosity of this process can be converted into a bound on
the fundamental Planck scale of the theory, $M_*$. The argument is very similar
to that used to bound the axion-nucleon coupling strength
\cite{Raffelt, axionpapers,BT,BBT}. The `standard model' of supernovae does an
exceptionally good job of predicting the duration and shape of the neutrino
pulse from SN1987A. Any mechanism which leads to significant energy-loss from
the core of the supernova immediately after bounce will produce a very different
neutrino-pulse shape, and so will destroy this agreement as demonstrated
explicitly in the axion case by Burrows, Brinkmann, and Turner \cite{BBT}.
Raffelt has proposed a simple analytic criterion based on detailed supernova
simulations \cite{Raffelt}: if any energy-loss mechanism has an emissivity
greater than $10^{19}$ ergs g$^-1$s$^-1$ then it will remove sufficient energy from the
explosion to invalidate the current understanding of SNe II
neutrino signal.

\section{Supernovae and constraints on large extra dimensions}
The most restrictive limits on $M_*$ come from SN1987A energy-loss argument.
If large extra dimensions exist, the usual four dimensional graviton is
complemented by a tower of Kaluza-Klein (KK) states, corresponding to new phase
space in the bulk.   The KK gravitons interact with the strength of ordinary
gravitons and thus are not trapped in the SN core.  During the first few seconds
after collapse, the core contains neutrons, protons, electrons, neutrinos and
thermal photons. There are a number of processes in which higher-dimensional
gravitons can be produced. For the conditions that pertain in the core at this
time (temperatures $T \sim 30-70$ MeV, densities $\rho \sim (3-10) \times
10^{14}$ g cm$^{-3}$), the relevant processes are  nucleon-nucleon
brehmstrahlung, graviton production in photon fusion and  electron-positron
annihilation.

In SNe, nucleon and photon abundances are comparable (actually nucleons are
somewhat more abundant). In the following we present the bounds derived by
various authors using nucleon-nucleon brehmstralung and in the next section we
give a detailed calculation for photon-photon annihilation to KK graviton process.

\subsection{Nucleon-Nucleon brehmstralung }
This is the dominant process relevant for the SN1987A where the temperature is
comparable to pion mass $m_\pi$ and so the strong interaction between nucleons
is unsuppressed. This process can be represented as
\begin{equation}
 N + N \to N + N + KK
 \end{equation}
where $N$ can be a neutron or a proton and $KK$ is a higher-dimensional
graviton.

The main uncertainty comes from the lack of precise knowledge of temperatures in
the core: values quoted in the
literature range from 30 to 70 MeV.  For  $T=30$ MeV and $\rho=3 \times 10^{14}$
g cm$^{-3}$, we list the results obtained by various authors. \\
Cullen and Perelstein \cite{Cullen:1999hc}
\begin{eqnarray}
n=2,\,\,\hskip0.5cm \dot{\epsilon}=6.79 \times 10^{25} \times M_*^{-4} \,erg\, g^{-1}\, s^{-1},\,\,\hskip0.5cm M_* &\gtrsim& 50 \hbox{TeV};\\
n=3,\,\,\hskip0.5cm \dot{\epsilon}=1.12 \times 10^{22} \times M_*^{-5} \,erg\, g^{-1}\, s^{-1},\,\,\hskip0.5cm M_* &\gtrsim&\,\,4 \hbox{TeV}.
\end{eqnarray}
Barger, Han, Kao and Zhang \cite{BHKZ}
\begin{eqnarray}
n=2,\,\,\hskip0.5cm \dot{\epsilon}=6.7\,\, \times 10^{25} \times M_*^{-4} \,erg\, g^{-1}\, s^{-1},\,\,\hskip0.5cm M_* &\gtrsim& 51 \hbox{TeV};\\
n=3,\,\,\hskip0.5cm \dot{\epsilon}=6.3\,\, \times 10^{21} \times M_*^{-5} \,erg\, g^{-1}\, s^{-1},\,\,\hskip0.5cm M_* &\gtrsim&  3.6 \hbox{TeV}.
\end{eqnarray}
Hanhart \textit{et. al.} \cite{Hanhart:2001er, Hanhart:2001fx}
\begin{eqnarray}
n=2,\,\,\hskip0.5cm \dot{\epsilon}=9.24 \times 10^{24} \times M_*^{-4} \,erg\, g^{-1}\, s^{-1},\,\,\hskip0.5cm M_* &\gtrsim& 31  \hbox{TeV};\\
n=3,\,\,\hskip0.5cm \dot{\epsilon}=1.57 \times 10^{21} \times M_*^{-5} \,erg\, g^{-1}\, s^{-1},\,\,\hskip0.5cm M_* &\gtrsim& \,\, 2.75 \hbox{TeV}.
\end{eqnarray}
Hannestad and Raffelt \cite{Hannestad:2001jv, Hannestad:2003yd}
\begin{eqnarray}
n=2,\,\,\hskip0.5cm \dot{\epsilon}=4.98 \times 10^{26} \times M_*^{-4} \,erg\, g^{-1}\, s^{-1},\,\,\hskip0.5cm M_* &\gtrsim& 84 \hbox{TeV};\\
n=3,\,\,\hskip0.5cm \dot{\epsilon}=1.68 \times 10^{23} \times M_*^{-5} \,erg\, g^{-1}\, s^{-1},\,\,\hskip0.5cm M_* &\gtrsim& \,\, 7 \hbox{TeV}.
\end{eqnarray}
\section{ Graviton production through photon fusion and energy loss rate}
Our aim is to study the energy loss mechanism of SN1987A by graviton emission by
photon-photon annihilation in the ADD framework. For this we need to compute the
cross-section for the relevant process. Here we present the general formalism
for calculating the cross-section \cite{QFT} for two particle initial state.
The scattering cross section is given by
\begin{eqnarray}
\sigma&=& \frac{1}{\upsilon_{rel}} \frac{1}{4 E_1 E_2} \int \prod_f \frac{d^3
p_f}{(2 \pi)^3 2 E_f}  \nonumber \\
&& \times (2 \pi)^4 \delta^4 \left( \sum_i p_i-\sum_f p_f\right)
|{\cal{M}}_{fi}|^2
\label{crosssec}
\end{eqnarray}
with
\begin{equation}
\upsilon_{rel}=\frac{\sqrt{(p_1\cdot p_2)^2-m_1^2 m_2^2}}{E_1 E_2},
\end{equation}
where $p_i$ and $E_i$ being the 3-momenta and the energies of the initial
particles whose masses are $m_1$ and $m_2$; $p_f$ and $E_f$ are the 3-momenta
and the energies of the final particles and ${\cal{M}}_{fi}$ is the Feynman
amplitude for the process.

For a general reaction of the kind $a+b \to c$, in the center-of-mass frame, the
expression (\ref{crosssec}) takes the form
\begin{equation}
\sigma= \frac{1}{64 \pi^2 E_1 E_2 \upsilon_{rel}} \int \frac{d^3 p'}{E'}
\delta(E_1+E_2-E') |{\cal{M}}|^2.
\label{cross}
\end{equation}
We use the center-of-mass frame, where we use the following notions.
\begin{eqnarray}
\sqrt{s}=E_1+E_2,\label{cmenergy}\\
E_1 E_2 \upsilon_{rel}=\mathbf{p} \sqrt{s},\label{relvel}
\end{eqnarray}
where $\mathbf{p}=\mathbf{p}_1 +\mathbf{p}_2$.

Next, we  focus on  the energy loss due to KK gravitons escaping into the extra
dimensions. The  energy loss as per unit time per unit mass of SN in terms of
the cross-section $\sigma_{a+b \to c}$, is given by \cite{Kolb}
\begin{equation}
\dot{\epsilon}_{a + b \to c.} = \frac{\langle n_a n_b
\sigma_{(a+b \to c)} v_{rel} E_{c} \rangle}{\rho}
\label{emrate}
\end{equation}
where the brackets indicate thermal average, $n_{a,b}$ are the number densities
for a, b and $\rho$ is the mass density and $E_c$ is the energy of the particle
c.

We  calculate the cross section using  the helicity method \cite{Jacob:at}-\cite{Ballestrero:1994jn}.
We follow the conventions and Feynman rules derived in \cite{HLZ}. In the
helicity method, it is more convenient to work with polarizations explicitly.
Thus, the polarization vectors \cite{Gleisberg:2003ue}
 of a massive graviton are
\begin{eqnarray}
e_{\mu\nu}^{\pm 2}&=&2\epsilon_{\mu}^{\pm}\epsilon_{\nu}^{\pm}
\ , \nonumber\\
e_{\mu\nu}^{\pm 1}&=&\sqrt{2}\, (\epsilon_{\mu}^{\pm}\epsilon_{\nu}^{0}+
\epsilon_{\mu}^{0}\epsilon_{\nu}^{\pm})\ ,\nonumber\\
e_{\mu\nu}^{0}&=&\sqrt{\frac{2}{3}}\, (\epsilon_{\mu}^{+}
\epsilon_{\nu}^{-}+
\epsilon_{\mu}^{-}\epsilon_{\nu}^{+}-
2\epsilon_{\mu}^{0}\epsilon_{\nu}^{0})\ . \nonumber
\end{eqnarray}
Here $\epsilon_{\mu}^{\pm}$ and $\epsilon_{\mu}^{0}$ are the
polarization vectors of a massive gauge boson; for a massive vector
boson with momentum $p^{\mu}=(E,0,0,p)$ and mass $m$,
\begin{eqnarray}
 \epsilon^+_{\mu}(p)&=&\frac{1}{\sqrt{2}}(0,1,i,0)\ ,   \\
 \epsilon^-_{\mu}(p)&=&\frac{1}{\sqrt{2}}(0,-1,i,0)\ ,   \\
 \epsilon^0_{\mu}(p)&=&\frac{1}{m}(p,0,0,-E)\ .
\end{eqnarray}
The graviton polarization vectors satisfy the normalization and
polarization sum conditions
\begin{eqnarray}
e^{s\,\mu\nu}e^{s'\, *}_{\mu\nu} &=& 4\delta^{s s'}\ ,\\
\sum_s e^s_{\mu\nu}e^{s\, *}_{\rho\sigma} &=& B_{\mu\nu\, \rho\sigma}\ ,
\end{eqnarray}
where $B_{\mu\nu\, \rho \sigma}$ is given by
\begin{eqnarray}
B_{\mu\nu\, \rho\sigma}(k) &=& 2\left(\eta_{\mu\rho}-\frac{k_\mu
k_\rho}{m_{\vec{n}}^2}\right) \left(\eta_{\nu\sigma}-\frac{k_\nu
k_\sigma}{m_{\vec{n}}^2}\right) \nonumber\\
&&+2\left(\eta_{\mu\sigma}-\frac{k_\mu
k_\sigma}{m_{\vec{n}}^2}\right)
\left(\eta_{\nu\rho}-{k_\nu k_\rho\over m_{\vec{n}}^2}\right)\nonumber\\
&& - \frac{4}{3} \left(\eta_{\mu\nu}-\frac{k_\mu
k_\nu}{m_{\vec{n}}^2}\right) \left(\eta_{\rho\sigma}-\frac{k_\rho
k_\sigma}{m_{\vec{n}}^2}\right)\ .
\label{B}
\end{eqnarray}
The total squared amplitude, averaged over the initial polarizations $z$ and
summed over final states for the reaction $a^{h}(q_1)+b^{h'}(q_2) \to
c^{h''}(p)$, is given by
\begin{equation}
\frac{1}{z} \sum_{h, h', h''} \left| {\cal M} \left(
a^{h}(q_1)+b^{h'}(q_2) \to c^{h''}(p) \right) \right|^2
\end{equation}
where $h, h', h''$ are the helicities and $q_1, q_2, p$ are the momenta of
particles a, b, c respectively.

Photons are quite abundant in supernovae. Here we consider photon-photon
annihilation to KK graviton and the process 
is given by,
\begin{equation}
\gamma(k_1) + \gamma(k_2) \to KK(p).
\label{photogravy}
\end{equation}

The vertex function  for the process (\ref{photogravy})  is given by \cite{HLZ}
\begin{eqnarray}
X_{\mu\nu\alpha\beta} &=& \frac{i}{2M_4}\biggl[\eta_{\alpha\beta}
k_{1 \mu} k_{2 \nu}
-\eta_{\mu\alpha}k_{1\beta}k_{2\nu}-\eta_{\nu\beta}k_{1\mu}k_{2\alpha}
\nonumber\\
&&+\eta_{\mu\alpha}\eta_{\nu\beta}(k_1\cdot k_2)
-\frac{1}{2}\eta_{\mu\nu}\left( \eta_{\alpha\beta}(k_1\cdot k_2)
 -k_{1\beta}k_{2\alpha} \right)\nonumber\\
 &&+ m_n m_{n-m} ( \eta_{\mu\alpha}\eta_{\nu\beta}
-\frac{1}{2} \eta_{\mu\nu}\eta_{\alpha\beta} ) + \left(\alpha
\leftrightarrow \beta \right) \biggr] .
\end{eqnarray}
The momentum vectors for this reaction are
\begin{eqnarray}
p^{\mu}\equiv (m_n, 0, 0, p)\\
k_1^{\mu}\equiv (k_1, 0, 0, k_1)\\
k_2^{\mu}\equiv (k_2, 0, 0, k_2).
\end{eqnarray}
In helicity formalism the reaction (\ref{photogravy}) can happen in two ways
\begin{eqnarray}
\gamma^{\pm}(k_1)+\gamma^{\pm}(k_2) \to KK^{\pm2}(p)\label{two}\\
\gamma^{\pm}(k_1)+\gamma^{\mp}(k_2) \to KK^{0}(p)\label{zero}.
\end{eqnarray}
Next, consider these two reactions separately and find their corresponding
amplitudes. For the reaction described in (\ref{two}), the helicity amplitude
for the KK graviton emission by photon-photon annihilation is
\begin{eqnarray}
 \left| {\cal M} \left(
\gamma^{\pm}(q)+\gamma^{\pm}(q) \to KK^{\pm2}(p)
\right) \right| 
= X^{\mu\nu\alpha\beta} \epsilon_{\alpha}^{\pm}(k) \epsilon_{\beta}^{\pm}(q)
e_{\mu\nu}^{\pm 2\, *}(p).
\label{plustwo}
\end{eqnarray}
The polarization tensors for gravitons are calculated and they are,
\begin{eqnarray}
e^{\pm2}_{11}=+{\frac{1}{2}},  \\
e^{\pm2}_{12}=e^{\pm2}_{21}=\frac{i}{2},   \\
e^{\pm2}_{22}=-{\frac{1}{2}}.
\end{eqnarray}
The non-zero components of the vertex function are $$ X^{1111}, X^{1212},  X^{1221}, X^{2112}, 
X^{2121}, X^{2222}, -X^{2211}, -X^{1122}$$  Each of them equal to
\begin{equation}
\frac{-i\kappa}{2}k_1 \cdot k_2.
\label{xresult}
\end{equation}
where $k_1 \cdot k_2=m_{\vec{n}}^2/2$.

Substituting the various quantities that we have calculated above in 
equation (\ref{plustwo}), we get
\begin{eqnarray}
\left| {\cal M} \left(
\gamma^{\pm}(q)+\gamma^{\pm}(q) \to KK^{\pm2}(p)
\right) \right|&=& {\kappa m_{\vec{n}}^2}\over 2.
\end{eqnarray}

The helicity amplitude for the reaction (\ref{zero}) is,
\begin{eqnarray}
 \left| {\cal M} \left(
\gamma^{\pm}(q)+\gamma^{\mp}(q) \to KK^{0}(p)
\right) \right|       
= X^{\mu\nu\alpha\beta} \epsilon_{\alpha}^{\pm}(k) \epsilon_{\beta}^{\mp}(q)
e_{\mu\nu}^{0\, *}(p).
\label{pluszero}
\end{eqnarray}
The polarization tensors for gravitons are given by
\begin{eqnarray}
e^0_{11}=e^0_{22}=-\sqrt{\frac{2}{3}} \\
e^0_{12}=e^0_{21}=0.
\end{eqnarray}
The non-zero components of the vertex function are $ X^{1111} \,
,X^{2222},-X^{2211}$ and $-X^{1122}$ and are equal to (\ref{xresult}).

Substituting the various quantities that we have calculated above in 
equation (\ref{pluszero}), we get
\begin{eqnarray}
\left| {\cal M} \left(
\gamma^{\pm}(q)+\gamma^{\mp}(q) \to KK^{0}(p)
\right) \right|
&=& 0.
\end{eqnarray}
Thus the total squared amplitude, averaged over the initial three polarizations
and summed over final states, is
\begin{equation}
\frac{1}{3} \sum_{h, h', h''} \left| {\cal M} \left(
\gamma^{h}(k_1)+\gamma^{h'}(k_2) \to KK^{h''}(p) \right) \right|^2
 = {{\kappa^2 m_{\vec{n}}^4}\over 12}.
\end{equation}
Substituting this in (\ref{cross}) and using (\ref{cmenergy}) and
(\ref{relvel}), the cross-section for the process is obtained as
\begin{eqnarray}
\sigma =\ {\pi\kappa^2 \sqrt{s}\over 16}
\delta (m_{\vec n}-\sqrt{s})\ ,
\end{eqnarray}
where $s$ is the center of mass energy, and $m_{\vec n}$ the mass
of the KK state at level $\vec n$.

Since for large $R$ the KK
gravitons are very light, they may be copiously produced in
high energy processes. For real emission of the KK gravitons
from a SM field, the total cross-section can be written as
\begin{equation}
\sigma_{\rm tot}\ =\ \kappa^2\sum_{\vec n} \sigma({\vec n})\ ,
\end{equation}
where the dependence on the gravitational coupling is factored out.
 The mass separation of adjacent KK states,  ${\cal O}(1/R)$,
is usually much smaller than typical energies in a physical process, therefore
we can approximate the summation by an integration which can be performed using
KK state density function \cite{HLZ},
\begin{equation}
\rho(m_{\vec{n}})=\frac{R^n m_{\vec{n}}^{n-2}}{(4 \pi)^{n/2} \Gamma(n/2)}.
\end{equation}
The volume emissivity of a supernova  with a temperature $T$  through the
process under consideration is obtained by thermal-averaging over  the
Bose-Einstein distribution
\begin{eqnarray}
Q_\gamma\ &=&\ \int {2 d^3{\vec k}_1\over (2\pi)^3} {1\over e^{\omega_1/T}-1}
\int {2 d^3{\vec k}_2\over (2\pi)^3} {1\over e^{\omega_2/T}-1}  \nonumber \\
&& \times {s(\omega_1+\omega_2)\over 2\omega_1\omega_2}\sum_{\vec n}
\sigma_{\gamma\gamma\rightarrow kk}(s,m_{\vec n}),
\end{eqnarray}
where the summation is over all KK states, and the squared center  of mass
energy $s$ is related to the photon energies  $\omega_1$ and $\omega_2$ and the
angle between the two photon momenta $\theta_{\gamma\gamma}$ as follows:
\begin{equation}
s\ =\ 2 \omega_1\omega_2 (1-\cos\theta_{\gamma\gamma})\ .
\end{equation}
After carrying out the integrals and the summation over KK states, we find
\begin{equation}
Q_\gamma\ =\ {2^{n+3}
\Gamma({n\over2}+3)\Gamma({n\over2}+4)\zeta({n\over2}+3)\zeta({n\over2}+4)
\over (n+4)\pi^2}
{T^{n+7}\over M_S^{n+2}} \ ,
\end{equation}
where we have used $M_*^{n+2} R^n S_n= M_{pl}^2$ and
numerically, these  Riemann zeta-functions are close to 1. In this calculation,
we have neglected the plasma effect, through which the photons can have
different energy  dispersion relations from those of free particles.

We take the supernova core density $\simeq 10^{15}$ g cm$^{-3}$. Using 
(\ref{emrate}), we compute the energy loss rate for $n=2$ and $n=3$ extra
spatial dimensions and hence the lower limits on $M_*$ using the conservative upper limits on 
the energy-loss rate of SN \cite{Raffelt}
\begin{equation}
{\dot{\epsilon}}_{SN}^{}\sim 10^{19}\ {\rm erg\ g}^{-1}{\rm sec}^{-1}.
\end{equation}
The results are summarized below,
\begin{eqnarray}
n=2,\,\,\hspace{0.5cm} \dot{\epsilon}= 4.7 \times 10^{23} \times M_*^{-4}\, erg\, g^{-1}\, sec^{-1},\,\,\hspace{0.5cm} M_* &\gtrsim& 14.72\hbox{TeV},\\
n=3,\,\,\hspace{0.5cm} \dot{\epsilon}= 1.1 \times 10^{20} \times M_*^{-5}\, erg\, g^{-1}\, sec^{-1},\,\,\hspace{0.5cm} M_* &\gtrsim& \,\,1.62\hbox{TeV}.
\end{eqnarray}

We now combine the energy loss rate due to photon fusion process with that of the nucleon-nucleon brehmstralung and rederive the constraints as follows.
Cullen and Perelstein \cite{Cullen:1999hc}
\begin{eqnarray}
n=2,\,\,\hskip0.5cm \dot{\epsilon}=6.837 \times 10^{25} \times M_*^{-4} \,erg\, g^{-1}\, s^{-1},\,\,\hskip0.5cm M_* &\gtrsim& 50.13 \hbox{TeV};\\
n=3,\,\,\hskip0.5cm \dot{\epsilon}=1.131 \times 10^{22} \times M_*^{-5} \,erg\, g^{-1}\, s^{-1},\,\,\hskip0.5cm M_* &\gtrsim&\,\,4.08 \hbox{TeV}.
\end{eqnarray}
Barger, Han, Kao and Zhang \cite{BHKZ}
\begin{eqnarray}
n=2,\,\,\hskip0.5cm \dot{\epsilon}=6.747 \times 10^{25} \times M_*^{-4} \,erg\, g^{-1}\, s^{-1},\,\,\hskip0.5cm M_* &\gtrsim& 50.96 \hbox{TeV};\\
n=3,\,\,\hskip0.5cm \dot{\epsilon}=6.410 \times 10^{21} \times M_*^{-5} \,erg\, g^{-1}\, s^{-1},\,\,\hskip0.5cm M_* &\gtrsim& \,\,3.64 \hbox{TeV}.
\end{eqnarray}
Hanhart \textit{et. al.} \cite{Hanhart:2001er, Hanhart:2001fx}
\begin{eqnarray}
n=2,\,\,\hskip0.5cm \dot{\epsilon}=9.710 \times 10^{24} \times M_*^{-4} \,erg\, g^{-1}\, s^{-1},\,\,\hskip0.5cm M_* &\gtrsim& 31.39  \hbox{TeV};\\
n=3,\,\,\hskip0.5cm \dot{\epsilon}=1.680 \times 10^{21} \times M_*^{-5} \,erg\, g^{-1}\, s^{-1},\,\,\hskip0.5cm M_* &\gtrsim& \,\,2.79 \hbox{TeV}.
\end{eqnarray}
Hannestad and Raffelt \cite{Hannestad:2001jv, Hannestad:2003yd}
\begin{eqnarray}
n=2,\,\,\hskip0.5cm \dot{\epsilon}=4.985 \times 10^{26} \times M_*^{-4} \,erg\, g^{-1}\, s^{-1},\,\,\hskip0.5cm M_* &\gtrsim& 84.03 \hbox{TeV};\\
n=3,\,\,\hskip0.5cm \dot{\epsilon}=1.681 \times 10^{23} \times M_*^{-5} \,erg\, g^{-1}\, s^{-1},\,\,\hskip0.5cm M_* &\gtrsim& \,\,7.00 \hbox{TeV}.
\end{eqnarray}
As we expected, the energy loss rate due to nucleon-nucleon brehmstralung is 1 to 3 orders of magnitude more than that due to photon fusion process. Hence the combined bounds  add only the second decimal place to $M_*$. 

\section{{Conclusions}}
We have revisited the constraints on ADD model coming from photon annihilation into KK graviton in SN cores. 
For a conservative choice of the core parameters, we obtain the two photon process bounds on the fundamental Planck scale $M_* \gtrsim$  1.6 TeV. The energy loss rate due to nucleon-nucleon brehmstralung is 1 to 3 orders of magnitude more than that due to photon fusion process. Hence the combined bounds  add only the second decimal place to $M_*$. Thus the present study can strengthen the results which are available in the current literature for  the  graviton emission from SN1987A. Our results show that the above processes put a very strong constraints on models with large extra dimensions for the case of $n=3$. Notice that the plasmon effects are not considered in our calculations and will be done elsewhere.

\section{{Acknowledgments}}
We thank Dr. Prasanta Kumar Das for collaboration in a similar project. The first author acknowledges Dr. R Chenraj Jain, Wg. Cdr. K L Ganesh Sharma,  Prof. T S Sridhar and Mr. M S Santhosh  for the kind hospitality and great facilities while writing this paper.

\end{document}